\documentclass[iop,apj]{emulateapj}
\usepackage{graphicx}

\shorttitle{The Role of Structured Magnetic Fields on Constraining Properties of Transient Sources of UHECRs}
\shortauthors{Takami \& Murase}

\begin{document}


\title{The Role of Structured Magnetic Fields on Constraining Properties of Transient Sources of Ultra-high-energy Cosmic Rays}

\author{Hajime Takami\altaffilmark{1} and Kohta Murase\altaffilmark{2,3}}
\email[E-mail address: ]{takami@mpp.mpg.de}

\altaffiltext{1}{Max Planck Institute for Physics, F\"ohringer Ring 6, 80805 Munich, Germany}
\altaffiltext{2}{Center for Cosmology and AstroParticle Physics, The Ohio State University, 191 W. Woodruff Ave., Columbus, OH 43210, USA}
\altaffiltext{3}{Department of Physics, The Ohio State University, 191 W. Woodruff Ave., Columbus, OH 43210, USA}

\begin{abstract}
We study how the properties of transient sources of ultra-high-energy cosmic rays (UHECRs) can be accessed by exploiting UHECR experiments, taking into account the propagation of UHECRs in magnetic structures which the sources are embedded in, i.e., clusters of galaxies and filamentary structures. Adopting simplified analytical models, we demonstrate that the structured extragalactic magnetic fields (EGMFs) play crucial roles in unveiling the properties of the transient sources. These EGMFs unavoidably cause significant delay in the arrival time of UHECRs as well as the Galactic magnetic field, even if the strength of magnetic fields in voids is zero. Then, we show that, given good knowledge on the structured EGMFs, UHECR observations with high statistics above $10^{20}$~eV allow us to constrain the generation rate of transient UHECR sources and their energy input per burst, which can be compared with the rates and energy release of known astrophysical phenomena. We also demonstrate that identifying the energy dependence of the apparent number density of UHECR sources at the highest energies is crucial as a clue to such transient sources. Future UHECR experiments with extremely large exposure are required to reveal the nature of transient UHECR sources.
\end{abstract}

\maketitle

\section{Introduction} \label{introduction}

The origin of ultra-high-energy cosmic rays (UHECRs) has been a mystery for more than forty years. The highest energy cosmic rays ($\gtrsim 10^{19}$ eV) are usually thought to be extragalactic origin, and various kinds of astrophysical objects have been suggested as primary source candidates, including gamma-ray bursts (GRBs)~\citep[e.g.,][]{Waxman:1995vg,Vietri1995ApJ453p883,Murase:2006mm,Murase2008PRD78p023005}, newly born magnetars~\citep{Arons2003ApJ589p871,Murase2009PRD79p103001,Kotera2011PhRvD84p023002}, active galactic nuclei (AGN)~\citep[e.g.,][]{Biermann1987ApJ322p643,Takahara:1990he,Norman:1995aa,Farrar:2008ex,Dermer2009NJPh11p5016,Pe'er2009PRD80p123018,Takami2011APh34p749,2011arXiv1107.5576M}, and structure formation shocks~\citep[e.g.,][]{Norman:1995aa,Kang1996ApJ456p422,Inoue:2007kn}. Theoretically, UHECR sources are expected to be powerful enough. For cosmic-ray accelerators associated with an outflow, the Hillas condition \citep{Hillas1984ARAA22p425} can be rewritten in terms of the isotropic luminosity $L$ as~\cite[e.g.,][]{bla00,Waxman:2003uj,Farrar:2008ex,lw09} 
\begin{equation}
L_B \equiv \epsilon_B L \gtrsim 2 \times 10^{45}~\frac{\Gamma^2 {E_{20}}^2}{Z^2 \beta} ~~{\rm erg}~{\rm s}^{-1}, 
\end{equation}
where $\epsilon_B$, $Z$, $\Gamma$, $\beta$ and $E_{20} = E / 10^{20}$~eV are a fraction of magnetic luminosity to the total luminosity, the nuclear mass number of cosmic rays, the bulk Lorentz factor of the outflow, the velocity of a shock or wave in the production region in the unit of speed of light and the energy of cosmic rays, respectively. Among known candidates, few steady sources such as Fanaroff-Riley (FR) II galaxies seem to satisfy this condition in local Universe for $Z = 1$, which is inconsistent with the observed anisotropy as long as UHECRs are protons~\citep[e.g.,][]{Takami2009Aph30p306}. Also, Zaw et al. (2009) argued that the power of AGN correlating with detected UHECRs seems insufficient to produce UHECR protons. The above luminosity requirement can be satisfied, however, if UHECRs are generated by powerful transient phenomena like AGN flares, GRBs and newly born magnetars even if they are protons~\citep[e.g.,][]{Farrar:2008ex,Dermer2009NJPh11p5016,lw09}.

The other possible astrophysical solution is to consider that heavy nuclei dominate over protons, where the required luminosity is reduced by $Z^2$ and therefore more objects are allowed to be UHECR sources. Indeed, the heavy-ion-dominated composition has been implied by recent results of the Pierre Auger Observatory (PAO)~\citep{Abraham2010PRL104p091101}. If this is the case, only a few nearby radio galaxies or even a single AGN such as Cen A may contribute to the observed UHECR flux~\citep[e.g.,][]{gor+08}.  Other sources, including radio-quiet AGN~\citep{Pe'er2009PRD80p123018} and GRBs~\citep{Murase2008PRD78p023005,Wang2008ApJ677p432}, are also viable. The absence of anisotropy at $\sim {10}^{20}~{\rm eV}/Z$ may imply high abundance of nuclei~\citep{lw09,abr11} even at thelower energies, the origin of which is unclear. On the other hand, the PAO data on the fluctuation of $X_{\rm max}$ seem difficult to be reconciled with the $X_{\rm max}$ distribution of the same data~\citep{anc11}, and proton composition may be possible with a different estimator of primary composition~\citep{ww11}.  Also, the High Resolution Fly's Eye (HiRes) has claimed proton-dominated composition even above $10^{19}$~eV~\citep{Abbasi2010PRL104p161101}. There are different arguments and the UHECR composition has not been settled experimentally. Proton composition seems possible at present.

If UHECR sources are transient, that is, the source activity is shorter than the dispersion of the arrival time produced by cosmic magnetic fields during propagation, the direct identification of UHECR sources by UHECR observations is a more difficult task than that for steady sources due to the delay of the arrival time between UHECRs and other neutral messengers (photons, neutrinos and gravitational waves) emitted by the same source activity. Multi-messenger approaches are definitely powerful, but it is important to extract as much information as possible from UHECR observations as one of the messengers. For transient sources, relations between observed quantities and the properties of UHECR sources have been discussed, considering intervening cosmic magnetic fields~\citep[e.g.,][]{MiraldaEscude1996ApJ462L59,Waxman1996ApJ472L89,Murase2008ApJ690L14}. Any candidate of primary UHECR sources has to possess enough energy budget to reproduce the observed flux, which is the product of the energy input per activity and the rate of bursts or flares. The rate of bursts or flares is related to the apparent UHECR source number density, which can be determined from anisotropy in UHECR arrival distribution. The relation between the rate and the apparent source number density is less obvious, depending on the Galactic magnetic field (GMF) and poorly known extragalactic magnetic fields (EGMFs). The GMF unavoidably affects UHECRs arriving at the Earth. Considering the GMF allows us to evaluate the above relation and to give a constraint on candidates of transient UHECR sources through comparing between the rate inferred from UHECR observations and that of known transient phenomena~\citep{Murase2008ApJ690L14}.

Despite current uncertainty in EGMFs, it has been believed that the inhomogeneity of the EGMFs in the Universe is crucial for the propagation of UHECRs~\citep[e.g.,][]{Sigl2003PRD68p043002,Sigl2004PRD70p043007,Takami2006ApJ639p803,Das2008ApJ682p29,Kotera2008PRD77p123003}. The Universe indeed has structures, which consist of clusters of galaxies, filaments, sheets and voids. It has been suggested that magnetic fields in the structured regions were amplified via cosmological structure formation, and various numerical simulations of the structure formation have shown that the EGMF distribution follows the matter distribution \citep[e.g.,][]{Sigl2003PRD68p043002,Sigl2004PRD70p043007,Dolag2005JCAP01p009,Ryu2008Science320p909}. So, astrophysical objects including astrophysical UHECR sources are generally embedded in the structured regions. Thus, these magnetic structures also unavoidably affect the propagation of UHECRs. The structured EGMFs play essential roles in the time-delay and time-profile spread of UHECRs as well as their deflections.

In this paper, we study the roles of structured EGMFs on the propagation of UHECRs, and see how the properties of transient UHECR sources, e.g., the rate of UHECR bursts or flares $\rho_s$, can be constrained. In section~\ref{idea}, we describe basic relations between the UHECR burst rate and observational quantities that can be obtained by UHECR experiments. Possible signatures of transient UHECR sources that may be seen in the arrival distribution of UHECRs are also discussed. In section~\ref{prop}, we numerically calculate the propagation of UHECRs in the magnetized structures for various source positions, and evaluate the time spread due to the structured fields. Then, in section~\ref{results}, we discuss possible constraints on the UHECR burst rate and cosmic-ray energy input burst, taking into account large uncertainty in the void EGMF. We present discussions in section~\ref{conclusion} and summarize this study in section~\ref{summary}. Throughout the paper, the proton-composition is assumed, and $\Lambda$CDM cosmology with $H_0 = 71$ km s$^{-1}$ Mpc$^{-1}$, $\Omega_m = 0.3$ and $\Omega_{\Lambda} = 0.7$ is adopted.

\section{Basic Relations} \label{idea}

\subsection{Maximum distance of contributing sources} 

UHECRs above $5 \times {10}^{19}$~eV cannot avoid energy-loss due to photomeson production with cosmic microwave background (CMB) photons~\citep{Greisen1966PRL16p748,Zatsepin1966JETP4L78}. Hence, nearby UHECR sources within the so-called Greisen-Zatsepin-Kuz'min (GZK) radius mainly contribute to the observed flux, which is typically the energy-loss length of $\sim 100$~Mpc at $6 \times {10}^{19}$~eV.

\begin{figure}
\includegraphics[clip,width=\linewidth]{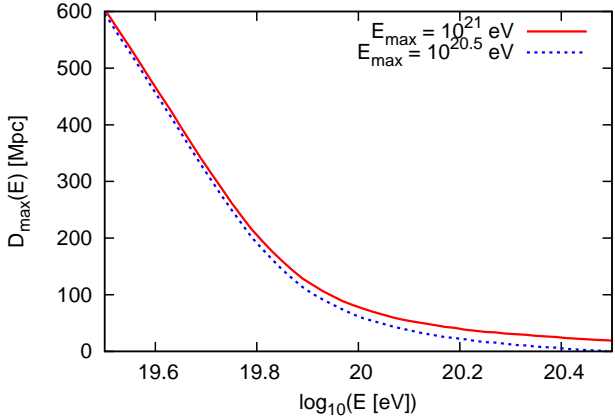}
\caption{Maximum distance of UHECR sources which can contribute to the flux observed at the Earth at the observed proton energy $E$ on the assumption of $E_{\rm max} = 10^{21}$ eV ({\it solid}) and $10^{20.5}$ eV ({\it dotted}). The maximum distance rapidly decreases due to photomeson production with CMB photons with increase of the proton energy. Note that $D_{\rm max}(E)$ is $200$~Mpc at $E = 6 \times 10^{19}$~eV and $75$~Mpc at $E = 10^{20}$~eV for $E_{\rm max} = 10^{21}$~eV.}
\label{fig:dmax}
\end{figure}

For practical purpose, we here introduce $D_{\rm max}(E)$ that is the maximum distance of sources of UHECRs observed with the energy $E$ at the Earth. This is defined as the distance within which sources are responsible for 99~\% of the total observed flux. This definition is more appropriate than what we used in \cite{Murase2008ApJ690L14}, where the energy-loss length of the photomeson production is simply adopted, because sources outside the energy-loss length can still significantly contribute to the total observed flux due to a volume effect. We estimate $D_{\rm max}(E)$ through a backtracking method of the propagation of UHECRs with a continuous energy-loss approximation on the photomeson production and Bethe-Heitler pair creation with CMB photons~\citep{Takami2006ApJ639p803}. Figure \ref{fig:dmax} shows $D_{\rm max}(E)$ calculated on the assmption that UHECR emissivity is proportional to ${(1 + z)}^3$ and the spectral index of -2.6, where $z$ is cosmological redshift. Note that $D_{\rm max}(E)$ is not sensitive on these assumptions since it is essentially determined by the energy-loss of the photomeson production. One sees that $D_{\rm max}(E)$ rapidly decreases owing to the photomeson production with CMB photons with increasing the energy of protons. $D_{\rm max}(E)$ is typically 200 Mpc and 75 Mpc at $E = 6 \times 10^{19}$ eV and $10^{20}$ eV for $E_{\rm max} = 10^{21}$ eV, respectively.

\subsection{Deflection and time profile spread} \label{def_td}

Trajectories of UHECRs are deflected in magnetized extragalactic space, whose magnetic fields are still uncertain \citep[e.g.,][for a review]{Kronberg1994RepProgPhys57p325}. Assuming that EGMFs are incoherent ($\lambda \ll D$), a typical deflection angle of UHECRs during propagation in extragalactic space is estimated to be \citep[][]{Waxman1996ApJ472L89}
\begin{eqnarray} 
\theta (E, D) &\approx& \frac{1}{r_{\rm L}} \left(\frac{2 D \lambda}{9} \right)^{1/2} \nonumber \\
&\simeq& 2^{\circ}~{B_{-9}} {\lambda_0}^{1/2} {E_{20}}^{-1} {D_{75}}^{1/2} \nonumber \\
&\lesssim& 20^{\circ} {E_{20}}^{-1} {D_{75}}^{1/2}, 
\label{eq:theta}
\end{eqnarray}
where $D_{75} = D / 75$ Mpc and $r_{\rm L}=E/eB$ are the propagation distance of UHECRs and the Larmor radius of the UHECRs for the characteristic EGMF strength of $B_{-9} = B / 10^{-9}$ G and the correlation length of $\lambda_0 = \lambda / 10^0$ Mpc, respectively. $e$ is the electron charge magnitude. The last inequality is obtained from an upper limit on the averaged EGMF from Faraday rotation measurements of distant quasars, $B {\lambda}^{1/2} \lesssim (10 {\rm nG}) (1 {\rm Mpc})^{1/2}$~\citep{Ryu1998AA335p19,Blasi1999ApJ514L79}. Similar values have been derived from CMB and matter power spectra for the primordial magnetic field at present \citep[e.g.,][]{Jedamzik2000PRL85p700,Yamazaki2010PRD81p023008}. On the other hand, several plausible GMF models predict deflection angles of $\lesssim 5^{\circ}$ at $\sim 6 \times 10^{19}$ eV and smaller at higher energies in almost all the directions in the sky~\citep{Takami:2007kq,Takami2010ApJ724p1456}. Those deflections seem comparable to the the correlation angle between UHECRs and their source candidates which were originally reported by the PAO, $\sim 3^{\circ}$ above $\sim 6 \times 10^{19}$ eV~\citep{Abraham2007Sci318p938}. However, one should keep in mind that such angular scale should not always be interpreted as the typical deflection angle of UHECRs, i.e., the source candidates could be tracers of true sources, both of which are embedded in the large-scale structure which are spread with typical angular scale of $\sim 10^{\circ}$ in local Universe in the projected sky. A careful correlation analysis on the PAO data with galaxies in local Universe suggested the correlation with the large scale structure, implying the deflection angles of $\lesssim 15^{\circ}$ more conservatively~\citep{Takami2008JCAP06p031}.

UHECRs propagate along different paths in magnetized space, which produce the time profile spread of a UHECR burst. The spread is comparable with the time-delay $t_d$~\citep[][]{MiraldaEscude1996ApJ462L59}. The apparent duration of a UHECR burst is 
\begin{equation}
\sigma (E, D) \sim t_d \approx \frac{D \theta^2(E, D)}{4 c}.
\label{eq:tau}
\end{equation}
This expression depends on $D$. Using $D_{\rm max}(E)$, let us introduce the characteristic time spread by~\citep[e.g.,][]{Murase2008ApJ690L14}
\begin{eqnarray}
\tau(E) &=& \frac{3}{4 \pi {D_{\rm max}}^3(E)} 
\int_0^{D_{\rm max}(E)} dD \, 4 \pi D^2 \sigma (E, D) \nonumber \\
&\simeq& \frac{3}{5} \sigma(E, D_{\rm max}(E)). 
\label{eq:chartau}
\end{eqnarray}

In the above discussion, we have not specified EGMFs. The EGMFs consist of magnetic fields in structured regions, i.e., clusters of galaxies and filaments, and magnetic fields in voids. Clusters of galaxies typically possess $\sim 0.1-1~\mu {\rm G}$, whereas magnetic fields in filaments are expected to be weaker. Recent sophisticated numerical simulations suggested $\sim 10$~nG and $\sim 30$~nG as volume-averaged and root-mean-square components, respectively~\citep{Ryu2008Science320p909}. The EGMF in a void is much less understood. Upper limits on the effective EGMF have been obtained from the Faraday rotation measure as mentioned above, i.e., $B \lambda^{1/2} \lesssim 10^{-8}~{\rm G}~{\rm Mpc}^{1/2}$. A recent evaluation of the upper limit of the primordial magnetic field, which is believed to be the EGMF in voids, derived $B_{\rm v} < 2.5$ nG for $\lambda_{\rm v} = 1$ Mpc from the power spectra of CMB and matter~\citep{Yamazaki2010PRD81p023008}. On the other hand, the usage of the pair-halo/echo emission from TeV blazars may give lower bounds~\citep[][]{pla95,mur08b}, and $B_{\rm v} \gtrsim 10^{-18}-10^{-17}$~G have recently been suggested from {\it Fermi} data~\citep[e.g.,][]{Dolag2011ApJ727L4,Dermer2011ApJ733L21,Takahashi2011arXiv11033835}. If the strength of the void EGMF is close to these lower limits, only EGMFs in the structure regions play the role in the propagation of UHECRs. The effective magnetic field strength $B_{\rm eff}$ and its correlation length $\lambda_{\rm eff}$ can be estimated from, 
\begin{equation}
\theta^2(E, D) \sim f_{\rm c} \frac{2 D \lambda_{\rm c}}{9 {r_{\rm L,c}}^2} 
+ f_{\rm f} \frac{2 D \lambda_{\rm f}}{9 {r_{\rm L,f}}^2} 
+ f_{\rm v} \frac{2 D \lambda_{\rm v}}{9 {r_{\rm L,v}}^2} 
\equiv \frac{2 D \lambda_{\rm eff}}{9 {r_{\rm L,eff}}^2}, 
\label{eq:effb}
\end{equation}
where $f_{\rm x}$, $\lambda_{\rm x}$ and $r_{\rm L,x}$ for ${\rm x} = {\rm c}$, ${\rm f}$, ${\rm v}$ are the volume filling fraction of a structure x, the correlation length of a magnetic field and the Larmor radius of UHECRs in a structure x, respectively, and ${\rm x} = {\rm c}$, ${\rm f}$, ${\rm v}$ correspond to clusters of galaxies, filamentary structures and voids, respectively.

Here, let us assume that the EGMF in voids is negligibly weak for the UHECR propagation, focusing on effects of the structured EGMFs. Because of the small volume filling fraction of galaxy clusters ($f_{\rm c} \sim 10^{-4}$) in our model (see the next section) and several simulation results~\citep[][see also Figure 9 of \cite{Kotera2011arXiv11014256}]{Dolag2005JCAP01p009,Das2008ApJ682p29}, one may approximate that only UHECRs generated in clusters of galaxies are affected by a magnetic field of clusters. Assuming $f_{\rm f} \sim 0.01$, $B_{\rm f} = 10$ nG and $\lambda_{\rm f} = 100$ kpc~\citep{Ryu2008Science320p909}, an estimated value is $B_{\rm eff} {\lambda_{\rm eff}}^{1/2} \sim 0.3$ nG Mpc$^{1/2}$ because only the second term contributes if all UHECR sources are embedded in filamentary structures, or $B_{\rm eff} {\lambda_{\rm eff}}^{1/2} \sim 1.0$ nG Mpc$^{1/2}$ if all UHECR sources are embedded in clusters of galaxies on the assumptions of $B_{\rm c} = 0.3 \mu$G and $\lambda_{\rm c} = 100$ kpc.

In this section, for simplicity, the energy-loss of protons was neglected except for $D_{\rm max}(E)$ in discussing the analytical expressions. The energy-loss during propagation will be taken into account when numerical calculations are performed in section \ref{prop} and \ref{results}. 

\subsection{UHECR burst rate and apparent source density} \label{ns}

We call sources "transients" when the intrinsic duration of UHECR production at a source $\delta T$ is shorter than the characteristic time profile spread $\tau(E)$ at the observed energy of $E$. If the time dispersion is longer than the time scale of UHECR observations, we misperceive that UHECR bursts are steady sources, and therefore can define the "apparent" number density of UHECR sources $n_s(E)$. The source number density is related to the rate of UHECR bursts $\rho_s$ as~\citep*{MiraldaEscude1996ApJ462L59}
\begin{equation}
\rho_s \approx \frac{n_s(E)}{\tau (E)}. 
\label{eq:estrate}
\end{equation}
The source number density generally depends on UHECR energies, since the apparent duration is dependent on energies explicitly.

UHECRs observed at the Earth suffer from the GMF and EGMFs embedding their sources.The GMF typically has order of $\mu {\rm G}$ for a disk component, and it also has random and halo components. In principle, $\rho_s$ can be estimated by equation~(\ref{eq:estrate}) if the time spread by these fields and the EGMF in voids can be well estimated. However, the EGMF in voids is highly uncertain as discussed in the last subsection, but could contribute to the total time spread significantly because of large propagation distance compared to the size of Galactic space and the magnetic structures around sources. This uncertainty leads to a finite range of allowed values of $\rho_s$. Given inevitable contributions of the GMF and EGMFs embedding UHECR sources to the apparent duration, $\tau_{\rm min}(E)$, and the allowed maximal time spread including the contribution from the poorly known EGMF in voids, $\tau_{\rm max}(E)$, the rate of UHECR bursts is limited as~\citep{Murase2008ApJ690L14} 
\begin{equation}
\frac{n_s(E)}{\tau_{\rm max}(E)} \lesssim \rho_s \lesssim 
\frac{n_s(E)}{\tau_{\rm min}(E)}. 
\label{eq:limit}
\end{equation}
Here, $n_s(E)$ can be, in principle, estimated from anisotropy in the arrival distribution of UHECRs~\citep[e.g.,][]{Yoshiguchi2003ApJ586p1211,Takami2006ApJ639p803,Takami2009Aph30p306,Cuoco2009ApJ702p825}, assuming that the time spread is longer than the UHECR observation timescale.

However, one should keep in mind that equation~(\ref{eq:estrate}) is valid when each UHECR burst can be individually identified as a burst~\citep{Murase2008ApJ690L14}. If more than one bursts or flares occurring in an angular patch contribute to UHECRs observed in the same time-window, i.e., the time profiles of two independent UHECR bursts from the same direction (within the size of the angular patch) are overlapped at the Earth, equation~(\ref{eq:estrate}) cannot be used as it is. Therefore, one has to focus on UHECRs with higher energies to examine cases where $\tau(E)$ is shorter than the apparent time interval between bursts or flares occurring in the same angular patch, $\Delta T$. In reality, UHECRs have finite deviation angles due to cosmic magnetic fields, so UHECRs from a source arrives within a finite solid angle $\Delta \Omega = \pi \psi^2$ around the source, which can be regarded as the appropriate size of the finite angular patch. For a given $\rho_s$, the apparent time interval between bursts in the region of the sky with $\Delta \Omega$ is estimated to be
\begin{eqnarray}
\Delta T &\sim& \frac{3}{\Delta \Omega {D_{\rm max}(E)}^3 \rho_s} \nonumber \\
&\sim& 3 \times 10^5 {\psi_5}^{-2} {\rho_{s,0}}^{-1} 
\left( \frac{D_{\rm max}(E)}{75~{\rm Mpc}} \right)^{-3} ~{\rm yr}, 
\label{eq:deltat}
\end{eqnarray}
where $\psi_5 \equiv \psi / 5^{\circ}$ and $\rho_{s,0} = \rho_s / 10^0$ Gpc$^{-3}$ yr$^{-1}$. We take the typical positional correlation scale as $\psi$, and use $\psi \sim 5^{\circ}$ as a reference choice, which corresponds to $B_{\rm eff} {\lambda_{\rm eff}}^{1/2}  \lesssim 2$~nG~Mpc$^{1/2}$. This is reasonable, since this is consistent with the effective EGMFs estimated in the last subsection and a current upper limit of the void EGMF from a plausible cosmological model is 2.5 nG for $\lambda_{\rm v} = 1$ Mpc \citep{Yamazaki2010PRD81p023008}, but more conservative discussions with larger values of $\psi$ are also possible. Equation~(\ref{eq:deltat}) implies that a smaller $\Delta \Omega$ gives larger $\Delta T$, but $\Delta T$ should be limited to the burst/flare intermittence in a host galaxy, $\approx n_h/\rho_s$, where $n_h$ is the number density of host galaxies of UHECR sources. In other words, $\Delta \Omega$ smaller than the corresponding lower limit is meaningless, at which one host galaxy should exist in a volume with a solid angle $\Delta \Omega$ within $D_{\rm max}(E)$.

\begin{figure}
\includegraphics[clip,width=\linewidth]{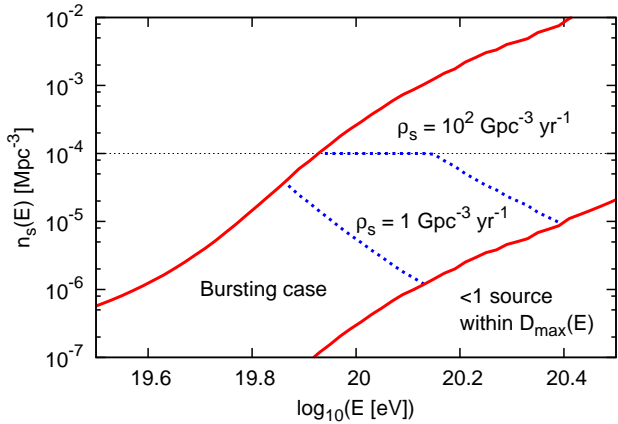}
\caption{The diagram of how transient UHECR sources are observed by UHECR experiments. In the region below the upper solid line each UHECR burst is observed with spatial and temporal separation, i.e., equation~(\ref{eq:estrate}) is valid. This case is called "bursting case" in this paper. Note that $\psi = 5^{\circ}$ is assumed. In the region below the lower solid line, there is no UHECR source within $D_{\rm max}(E)$. The relation between the apparent source number density $n_s(E)$ and $E$ is demonstrated ({\it dashed lines}) for $\rho_s = 1$~Gpc$^{-3}$~yr$^{-1}$ and $\rho_s = 10^2$~Gpc$^{-3}$~yr$^{-1}$ in the case where a EGMF in filamentary structures dominantly affects the propagation of UHECRs, i.e., $B_{\rm eff} {\lambda_{\rm eff}}^{1/2} \sim 0.3$~nG~Mpc$^{-3}$. The curve in the case of $\rho_s = 10^2$~Gpc$^{-3}$~yr$^{-1}$ is saturated at $n_h$, which is assumed to be $10^{-4}$~Mpc$^{-3}$.} 
\label{fig:nlim}
\end{figure}

We call the case "bursting case" that only one burst or flare contributes to arriving UHECRs at a time in a direction, i.e., $\tau(E) < \Delta T$. Then, the requirement $\tau(E) < \Delta T$ gives a sufficient condition to apply equation~(\ref{eq:estrate}), which leads to  
\begin{equation}
n_s (E) \lesssim 3 \times 10^{-4} {\psi_5}^{-2} 
\left( \frac{D_{\rm max}(E)}{75 {\rm Mpc}} \right)^{-3} ~{\rm Mpc}^{-3}, 
\label{eq:nslimit}
\end{equation}
with the usage of equation~(\ref{eq:estrate}). As demonstrated in Figure~\ref{fig:nlim}, the range of $n_s(E)$, in which equation~(\ref{eq:estrate}) can be applied, is extended to larger $n_s(E)$ at higher energies, because the smaller number of sources contributing to the observed flux decreases the probability that two bursts are temporally overlapped in a region of the sky (see also equation~(\ref{eq:deltat})). Thus, we especially focus on cases of $E \sim {10}^{20}$~eV to demonstrate constraints on $\rho_s$ in Sections~\ref{prop} and \ref{results}, although discussions are general for other $E$. Note that, although we here fix $\psi$ even at higher energies, smaller values of $\psi$ are expected there, so that the extension of the curve to higher energies would be more easily justified.

On the other hand, if $\tau(E) > \Delta T$, another UHECR burst may start to contribute before the end of the former UHECR burst is observed, and equation~(\ref{eq:deltat}) implies, 
\begin{equation}
\rho_s \gtrsim 3 \times 10^5 \frac{{\psi_5}^{-2}}{\tau(E)} 
\left( \frac{D_{\rm max}(E)}{75 {\rm Mpc}} \right)^{-3} 
~{\rm Gpc}^{-3}~{\rm yr}^{-1}. 
\label{eq:limitapp}
\end{equation}

Since $n_s(E)$ can be determined by the auto-correlation analysis, equation~(\ref{eq:estrate}) enables us to estimate $\rho_s$ from observational quantities, if $\tau(E)$ can be evaluated by EGMF simulations and observations. Importantly, $n_s(E)$ has the characteristic energy dependence, which is demonstrated in Figure~\ref{fig:nlim}. Here, the case that only the EGMF in filamentary structures affects the time spread of UHECR bursts is considered for demonstration, i.e., $B_{\rm eff} {\lambda_{\rm eff}}^{1/2} \sim 0.3$ nG Mpc$^{-3}$. Two representative cases for $\rho_s$ are shown, i.e., $\rho_s = 1$ Gpc$^{-3}$ yr$^{-1}$ and $10^2$ Gpc$^{-3}$ yr$^{-1}$. One sees that $n_s(E)$ changes by more than one order of magnitude when $E$ increases by the cubic root of ten. This means that anisotropy features are different among energies. Thus, observations of UHECRs above $10^{20}$ eV are crucial to identify this tendency clearly. Future UHECR experiments with large exposures may detect a large number of the highest-energy events and allow us to determine the dependence of $n_s(E)$.

Now, remember that, even if $\psi$ was small, $\Delta T$ would be limited by $n_h/\rho_s$. In other words, $n_s(E)$ must not be larger than $n_h$, and thus the curve of $n_s(E)$ may be saturated. This situation is also demonstrated, assuming $n_h = 10^{-4}$ Mpc$^{-3}$, which is comparable to the local number density of FR I galaxies~\citep{Padovani1990ApJ356p75}. In the case of $\rho_s = 10^2$ Gpc$^{-3}$ yr$^{-1}$, the curve of $n_s(E)$ is saturated and then becomes flat at low energies. Thus, the number density of host galaxies could also be estimated if such a saturated curve is seen. Note that $n_h$ does not depend on a given EGMF model, and gives a robust constraint on UHECR source population through relations between UHECR sources and their host galaxies. The relations were briefly discussed in \cite{2009arXiv0910.2765T}.

There is a region on the diagram, where less than one transient source contributes to the total flux observed at the Earth in the bursting case, especially at high energies. This is obviously expected from the fact that $D_{\rm max}(E)$ becomes smaller at higher energies. This region cannot be used to constrain UHECR sources because there is no UHECR source in the sky. The condition that less than one transient source contributes to the total flux is 
\begin{equation}
\frac{4 \pi}{3} {D_{\rm max}}^3(E) \rho_s \sigma(E, D_{\rm max}(E)) \lesssim 1. 
\end{equation}
Remembering $n_s(E) \simeq 3 \rho_s \sigma(E, D_{\rm max}(E)) / 5$, the region where less than one source exists within $D_{\rm max}(E)$ is determined by
\begin{equation}
n_s(E) \lesssim 3 \times 10^{-7} \left( \frac{D_{\rm max}(E)}{75~{\rm Mpc}} \right)^{-3} ~~{\rm Mpc}^{-3}. 
\end{equation}
This border line is also shown in Figure~\ref{fig:nlim}, below which the curves of $n_s(E)$ are truncated.  This region does not depend on the choice of $\psi$.  Just above the border line, there are only a few sources within $D_{\rm max}(E)$, where we can observe strong anisotropy from these nearby rare sources, which is strong evidence for the location of UHECR sources. In this viewpoint large UHECR experiments for UHECRs above $E > 10^{20}$~eV are required.

The energy dependence of $n_s(E)$ may originate from transient source scenarios, but it is not always direct evidence on transient UHECR sources, although the dependence is generally different between transient scenarios and steady scenarios. If all the UHECR sources are identical and steady, $n_s(E)$ is a constant up to $E = E_{\rm max}$. However, if steady sources have different $E_{\rm max}$ among sources, $n_s(E)$ could depend on $E$. For instance, \cite{Kachelriess:2005xh} demonstrated that steady UHECR sources with the power-law spectral slope of $\sim 2.0$ can reproduce the observed steep spectrum if a power-law distribution of $E_{\rm max}$ with the index of $\sim 1.7$ among sources is assumed. In this case $n_s(E)$ is expected to be proportional to $E^{-1.7}$. On the other hand, in the cases of transient sources demonstrated in Figure~\ref{fig:nlim}, the spectral index of the dependence of $n_s(E)$ on $E$ is much less than $-2.0$ above $10^{20}$~eV owing to the dependence of $D_{\rm max}(E)$ on $E$. Different $E_{\rm max}$ among sources is expected to lead to additional steepening of the dependence, although an identical $E_{\rm max}$ is assumed throughout this paper for simplicity. Therefore, transient cases has the steeper dependence on $E$, but careful discussions are required to distinguish transient cases from steady cases. Spiky spectral features in the spectrum of UHECRs which are specific in transient sources may also help support transient scenarios ~\citep{MiraldaEscude1996ApJ462L59}. 

We have shown that the dependence of $n_s(E)$ on $E$ provides us with a hint of transient sources. Although above discussions neglect the energy loss of UHECRs, the tendency that $n_s(E)$ decreases at high energies is a general feature. In order to estimate $n_s(E)$ at sufficiently high energies, large statistics beyond the GZK energy ($\sim 10^{20}$~eV) is required. Large statistics can also reduce the uncertainty of estimated $n_s(E)$ significantly \citep{Takami2007Aph28p529}. Future large UHECR experiments to target UHECRs above $10^{20}$~eV such as Extreme Universe Space Observatory (JEM-EUSO) \citep{Ebisuzaki2008NuPhS175p237} and the Northern site of the Pierre Auger Observatory \citep{Bluemer2010NJPh12c5001} will give important information to constrain the properties of transient UHECR sources. 

\section{Propagation of UHECRs in Magnetized Regions} \label{prop}

The positions of UHECR sources in structured regions depend on source candidates. Recent studies have shown that the star formation rate of galaxies tends to be larger in the outskirts of clusters and filaments than that inside clusters \citep[e.g.,][]{Porter2008MNRAS388p1152}. Since the explosion of massive stars is expected to happen in galaxies with high star formation rates, UHECR source candidates such as GRBs and newly born magnetars are more likely to be located in filaments or outskirts of clusters. On the other hand, powerful radio galaxies may be more centrally distributed~\citep[e.g.,][]{Lin2007ApJS170p71}. Here, two EGMFs, i.e., the EGMF in a filamentary structure and the EGMF in a cluster of galaxies, and three locations of UHECR sources are investigated: in a filamentary structure, at the center of a cluster of galaxies and 1 Mpc away from the center of a cluster of galaxies. In addition, the GMF is also considered as an unavoidable magnetic field for UHECRs arriving at the Earth. 

\subsection{Models and calculation method}

Here, a cluster of galaxies is modeled as a spherical structure with the radius of 3 Mpc. Since the local number density of clusters is $\sim 10^{-6}$ Mpc$^{-3}$ \citep{Mazure1996AA310p31}, the volume filling fraction of galaxy clusters in this model is $\sim 10^{-4}$, which is the referred value in section \ref{ns}.

The strength of a magnetic field in the cluster is assumed to scale with the distance from the center of the cluster as given by the flux-freezing condition with an thermal electron component modeled as a $\beta$-model following the treatment of \cite{DeMarco2006PhRvD73p043004}, 
\begin{equation}
B(r) = B_0 \left( 1 + \frac{r}{r_c} \right)^{-0.7}, 
\end{equation}
where $r_c = 378$ kpc and $B_0 = 1 \mu$G. The direction of the magnetic field is set to be turbulent with the Kolmogorov power spectrum with the maximum scale of $\lambda_{\rm c,max}$. The strength of the magnetic field averaged over this cluster model is $\sim 0.3 \mu$G.

\begin{figure}
\includegraphics[clip,width=\linewidth]{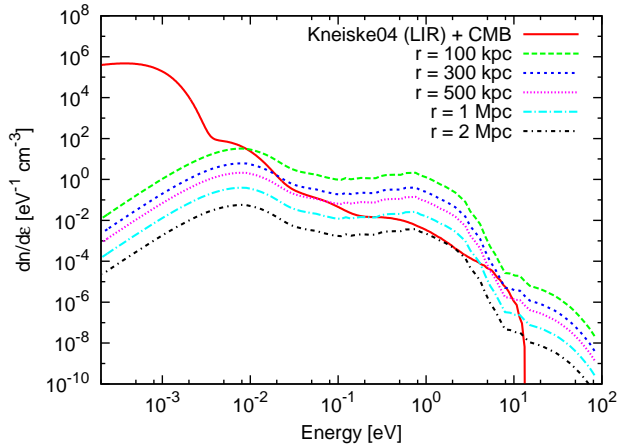}
\caption{Number densities of extragalactic background light including the CMB (the low infrared model of \cite{Kneiske2004AA413p807}) ({\it red}) and of background light in a cluster of galaxies at 100 kpc ({\it green}), 300 kpc ({\it blue}), 500 kpc ({\it magenta}), 1 Mpc ({\it light blue}), and 2 Mpc ({\it black}) away from the center of the cluster.}
\label{fig:bgd}
\end{figure}

An infrared background photon field in the cluster of galaxies is modeled as the superposition of the spectral energy distribution of 100 giant elliptical galaxies calculated by {\it GRASIL} \citep{Silva1998ApJ509p103} in addition to a low infrared model of extragalactic background light at $z = 0$ by \cite{Kneiske2004AA413p807}. The galaxies are assumed to be distributed following an analytical fitting formula of the gas distribution \citep{Rordorf2004APh22p167}, 
\begin{equation}
f(r) \propto 
\left[ 1 + \left( \frac{r}{r_1} \right)^2 \right]^{\alpha_1} 
\left[ 1 + \left( \frac{r}{r_2} \right)^2 \right]^{\alpha_2} 
\left[ 1 + \left( \frac{r}{r_3} \right)^2 \right]^{\alpha_3}, 
\end{equation}
where $r_1 = 10$ kpc, $r_2 = 250$ kpc, $r_3 = 1$ Mpc, $\alpha_1 = -0.51$, $\alpha_2 = -0.72$, and $\alpha_3 = -0.58$. Figure \ref{fig:bgd} shows the number densities of photons in the cluster of galaxies at several radius from the center.

As mentioned above, the two positions of a UHECR source in the cluster of galaxies are considered. In both cases, UHECRs are injected as a jet in the direction of an observer, i.e., in the radial direction, with the opening angle of 0.1 radian ($\sim 6^{\circ}$). The jet-like UHECR injection is motivated by many source candidates such as GRBs and AGN.

A filamentary structure is approximated to be a cylinder with the radius of 2 Mpc with the magnetic strength of 10 nG based on simulation results \citep{Ryu2008Science320p909}. Although the direction of a magnetic field is not clear yet, we assume a turbulent field with the Kolmogorov spectrum with the maximum scale of $\lambda_{\rm f,max}$. The photon field in the filament is assumed to be the same model of extragalactic background light as that used for the cluster. A source is located on the axis of the cylindrical filament and emits UHECRs toward a direction perpendicular to the axis with the jet opening angle of 0.1 radian. Although the direction of the jet with regard to the axis of the cylindrical filament depends on individual sources, this configuration is justified because it provides with a reasonable lower limit of the time-delay and resultant time spread of UHECRs. In addition, this allows us to avoid the dependence of results on the height of the cylinder, although it is assumed to be 25 Mpc.

Regarding the correlation length of the magnetic fields, the energy of turbulence is, in general, injected at large scale characterized by the driving scale of a turbulent magnetic field. The energy is transferred into smaller scale turbulence, and then is dissipated to thermal energy at small scale. In the intermediate scale, the energy spectrum of the turbulence can be well described by a power-law spectrum in many cases. Focusing on this region we model the EGMFs as Kolmogorov turbulence. However, these models do not consider magnetic fields at spatial scale larger than that where the power-law spectrum is valid. The ignorance of the larger scale magnetic fields let us underestimate the deflection angles of UHECRs. Since the spectral shape of the magnetic fields depends on their generation mechanism, scale and geometry, we avoid the uncertainty and simply employ Kolmogorov turbulence with the maximum scale so larger than the correlation lengths $\lambda_{\rm c}$ and $\lambda_{\rm f}$, as to reproduce the analytical estimation of UHECR deflection angles in the last subsection on average. We set $\lambda_{\rm c,max} = 400$ kpc and $\lambda_{\rm f,max} = 400$ kpc for the maximum scale.

In any case, once a UHECR is emitted from a source, the propagation of the UHECR is calculated by solving the equation of motion in each magnetic structure taking energy-loss processes by interactions with ambient photon fields, i.e., photomeson production and Bethe-Heitler pair creation, into account. An event generator SOPHIA is adopted to simulate the photomeson production \citep{Mucke1999CPC124p290}, and the energy-loss rate of protons by the Bethe-Heitler pair creation is estimated by an analytical fitting formula in \cite{Chodorowski1992ApJ400p181}. When the UHECR reaches the boundary of the structures, we record the time-delay and deflection angle of the UHECR, and obtain these distributions from the propagation results of many UHECRs.

The time-delay and deflection angles of UHECRs in the GMF is estimated by using a backtracking method \citep[e.g.,][]{Takami:2007kq,Takami2010ApJ724p1456}. For the GMF model, the BS-S model used in \cite{Takami2010ApJ724p1456}, which was originally proposed by \cite{AlvarezMuniz:2001vf}, is adopted. Since the GMF is distributed similarly to the Galactic arm, the time-delay of UHECRs depends on the arrival directions of UHECRs. So, we adopt a value averaged over the whole sky as a typical value of time spread. All the energy-loss processes can be neglected because the propagation path length of protons is much smaller than their attenuation length at energies we are interested in.

\subsection{Time-delay and deflections}
 
\begin{figure}
\includegraphics[clip,width=\linewidth]{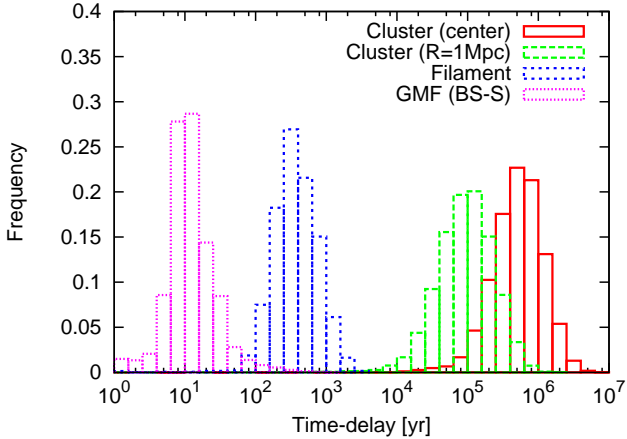}
\caption{Normalized time-delay distributions of protons with energies of $10^{20}$ eV produced during propagation in the GMF ({\it magenta}) and extragalactic magnetic structures around a source, i.e., a cluster of galaxies when the source is located at the center ({\it red}), a cluster of galaxies if there is the source at 1 Mpc away from the center ({\it green}) and a filamentary structure ({\it blue}). Note that the $10^{20}$ eV is the energy of protons at the local frame of the magnetic structures. It is the energy of protons at UHECR generation in the former three cases, while it is the energy when UHECRs penetrate into the Galactic space, i.e., observed energy, in the latter case.}
\label{fig:td200}
\end{figure}

\begin{figure}
\includegraphics[clip,width=\linewidth]{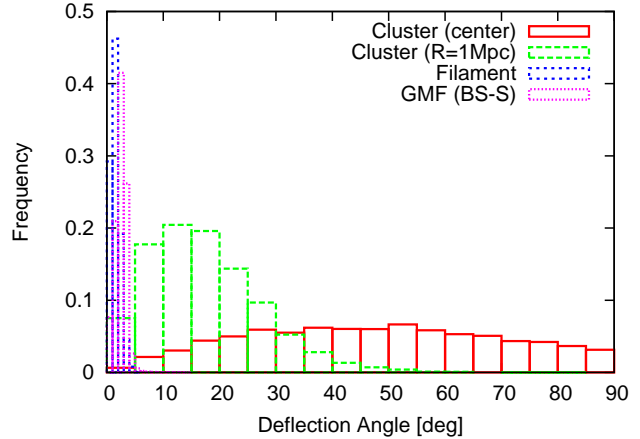}
\caption{Normalized deflection-angle distributions of protons with energies of $10^{20}$ eV produced during propagation in the different magnetized regions corresponding to the structures considered in Figure \ref{fig:td200}. The definitions of the labels and the energy of $10^{20}$ eV are the same as in Figure \ref{fig:td200}.}
\label{fig:def200}
\end{figure}

Figure~\ref{fig:td200} shows the distribution of the time-delay of protons with the energy of $10^{20}$ eV in the different magnetic structures. $\sigma(E) \sim \bar{t}_{\rm d}(E)$ is confirmed in all the cases. The deviation of the time profile of a UHECR burst produced by these magnetic structures are $\sigma_{\rm G}(E) \sim 10^{1.5}$ yr, $\sigma_{\rm fil}(E) \sim 10^{2.5}$ yr, $\sigma_{\rm c1}(E) \sim 10^{5}$ yr and $\sigma_{\rm cc}(E) \sim 10^{5.8}$ yr at $E = 10^{20}$ eV, respectively. Here $\sigma_{\rm G}(E)$ and $\sigma_{\rm fil}(E)$ are the time spread produced by the GMF and the EGMF in a filamentary structure, respectively. Both $\sigma_{\rm cc}(E)$ and $\sigma_{\rm c1}(E)$ are time spreads produced by magnetic field in a cluster of galaxies, but the former is the case when a UHECR source is located at the center of the cluster, while the latter is the case when a UHECR source is located at 1 Mpc away from the center of the cluster. $\sigma_{\rm fil}(E)$, $\sigma_{\rm c1}(E)$ and $\sigma_{\rm cc}(E)$ are proportional to $E^{-2}$ because the turbulent fields are assumed, which were able to be confirmed by the numerical simulations.

Here, let us check the validity of $\psi \sim 5^{\circ}$. Figure \ref{fig:def200} shows the distribution of the deflection angles of protons with the energy of $10^{20}$ eV in the different magnetic structures. The deflection of the protons by the GMF and magnetic field in a filamentary structure satisfy $\psi \lesssim 5^{\circ}$. On the other hand, a cluster of galaxies produces the deflection of protons larger than $5^{\circ}$. However, in this case, the sky region where UHECRs from the source are occupied is determined not by the deflection angles but by a viewing angle of the magnetic structure; i.e., $\psi \lesssim 5^{\circ}$ is satisfied if $D \gtrsim 30$ Mpc. Thus, only the Virgo cluster ($\sim 16$ Mpc) does not satisfy this requirement. If the Virgo cluster does not have UHECR sources, $\psi \sim 5^{\circ}$ is still justified. Otherwise, it is sufficient to continue discussions that a larger $\psi \lesssim 10^{\circ}$ is adopted. In the latter case, the validity of equation~(\ref{eq:limit}) is a bit changed as shown in equation~(\ref{eq:nslimit}). We use $\psi = 5^{\circ}$ to simply compare the case of clusters of galaxies with the other cases in this paper. 

\section{Implications for properties of UHECR bursts} \label{results}

For UHECRs observed at energies $E$, while the time spread by the GMF should be estimated at the observed energy $E$, that by structured EGMFs embedding their sources should be considered by using $E_g(E,D)$, which is the energy of UHECRs at generation, because UHECRs lose their energies during propagation in intergalactic space. This is specially important when $E$ is higher than the threshold of the photomeson production. Assuming that UHECR sources are uniformly distributed in local Universe, we introduce characteristic time spread in the structured EGMFs as 
\begin{equation}
\tau_{\rm x}(E) = \frac{3}{4 \pi {D_{\rm max}}^3(E)} 
\int_0^{D_{\rm max}(E)} dD~4 \pi D^2 \sigma_{\rm x}(E_g(E,D),D), 
\end{equation}
where x $=$ fil, cc, c1. Here $E_g(E,D)$ is calculated by the backtracking method used for the calculations of $D_{\rm max}(E)$. We adopt the characteristic time profile spread $\tau_{\rm x}(E)$ to constrain $\rho_s$ below. Since $E$ is lower than $E_g(E,D)$ for any $E$ and $D$ because of energy-loss, $\sigma_{\rm x}(E_{\rm max})$ is generally smaller than $\tau_{\rm x}(E)$. Therefore, $\sigma_{\rm x}(E_{\rm max})$ instead of $\tau_{\rm x}(E)$ gives a more conservative constraint.

The minimum of the characteristic time spread $\tau_{\rm min}(E)$ is estimated for the three cases of source locations. If the sources are located in filamentary structures, the characteristic time spread is estimated as $\tau_{\rm min}(E) \approx {\rm max}(\sigma_{\rm G}(E), \tau_{\rm fil}(E)) \approx \tau_{\rm fil}(E)$, where $\tau_{\rm fil}(E) \sim 57$ yr at $E = 10^{20}$ eV. For the remaining two cases where UHECR sources are embedded in clusters of galaxies, the time spread produced by the clusters is much longer than that produced by the GMF, i.e., $\tau_{\rm min}(E) \approx \tau_{\rm cc}(E) \sim 9 \times 10^4$ yr and $\tau_{\rm c1}(E) \sim 2 \times 10^4$ yr at $E = 10^{20}$ eV, respectively. These values constrain $\rho_s$ from an upper side in the bursting case.

As mentioned before, the EGMF in voids may mainly contribute to the total time spread of UHECR bursts. Following equations~(\ref{eq:tau}) and (\ref{eq:chartau}), and $\psi = 5^{\circ}$, the characteristic time spread by the void EGMF is limited as 
\begin{eqnarray}
\tau_{\rm v} (E) &\simeq& \frac{3}{5} \sigma_{\rm v} (E, D_{\rm max}(E)) \nonumber \\
&\lesssim& 1 \times 10^6 {E_{20}}^{-2} \left( \frac{D_{\rm max}(E)}{75~{\rm Mpc}} \right)^2 ~ {\rm yr} \equiv \tau_{\rm v,max}(E). 
\end{eqnarray}
Here, $\tau_{\rm v}(E)$ dominantly determines $\tau_{\rm max}(E)$ in all the cases, when $\tau_{\rm v,max}(E)$ is larger than $\tau_{\rm fil}(E)$, $\tau_{\rm c1}(E)$, and $\tau_{\rm cc}(E)$.

\begin{figure}
\includegraphics[clip,width=\linewidth]{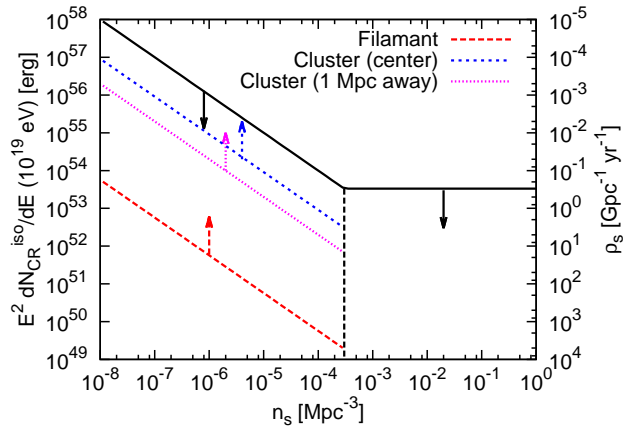}
\caption{Possible constraints on the (differential) cosmic-ray energy input at $10^{19}$ eV per burst $E^2 (dN_{\rm CR}^{\rm iso}/dE)$ and the rate of bursts or flares $\rho_s$ as functions of the apparent source number density $n_s(E)$. The vertical dashed line distinguishes between the bursting case and the other. In the bursting case (the left side of the line) these quantities are limited from both sides by equation~(\ref{eq:limit}). The black solid line shows the left inequality of equation~(\ref{eq:limit}), in which $\tau_{\rm max}(E)$ is dominated by the effective time spread of UHECRs by the void EGMF $\tau_{\rm v}(E)$. The three color lines correspond to the right inequality of equation~(\ref{eq:limit}) for each magnetic environment around UHECR sources, i.e., filamentary structures ({\it red}), the case where sources exist at the center of clusters of galaxies ({\it blue}), and the case where sources are located at 1 Mpc away from the center ({\it magenta}). On the other hand, at the right side of the dashed line, these quantities are limited only from one side by equation~(\ref{eq:limitapp}). Note that $E = 10^{20}$ eV, $E_{\rm max} = 10^{21}$ eV and $\psi = 5^{\circ}$ are assumed.}
\label{fig:einput}
\end{figure}

Then, we can constrain the rate of UHECR bursts following equation~(\ref{eq:limit}) for the bursting case, and otherwise $\rho_s$ is constrained by equation~(\ref{eq:limitapp}), into which $\tau_{\rm max}(E)$ is substituted instead of $\tau(E)$. Emissivity of UHECRs at $10^{19}$ eV required to reproduce the observed UHECR flux is $\tilde{\mathcal{L}}_{\rm CR} \sim 10^{44}$ erg Mpc$^{-3}$ yr$^{-1}$ \citep{Waxman:1998yy,Berezinsky:2002nc,Murase2008ApJ690L14}. This emissivity can be converted into the requirement of cosmic ray energy input per burst by using $\rho_s$, $\tilde{\mathcal{E}}_{\rm CR}^{\rm iso} = E^2 (dN_{\rm CR}^{\rm iso} / dE) = \tilde{\mathcal{L}}_{\rm CR}/ \rho_s$. Figure \ref{fig:einput} summarizes constraints on $\rho_s$ and $\tilde{\mathcal{E}}_{\rm CR}^{\rm iso}$ of transient UHECR sources as functions of $n_s(E)$. The black solid line represents a constraint of $\rho_s$ derived from $\tau_{\rm v,max}(E)$. The three inclined lines show constraints on $\rho_s$ or $\tilde{\mathcal{E}}_{\rm CR}^{\rm iso}$. The vertical black dashed line is $n_s(E) = 3 \times 10^{-4} {\psi_5}^{-2} (D_{\rm max}(E) / 75 {\rm Mpc})^{-3}$ Mpc$^{-3}$ for $E = 10^{20}$ eV (see equation~(\ref{eq:nslimit})), which divide the bursting case from the other.

\section{Discussion} \label{conclusion}

Although we focused on transient UHECR sources in this work, one should keep in mind that the information on $n_s (E)$ itself is relevant, whether sources are steady or transient. Since actual $n_s(E)$ must be smaller than $n_h$ at any energy $E$ (unless multiple images caused by sufficiently strong EGMFs lead to difficulty in determination of plausible $n_s (E)$), comparing $n_s (E)$ with the number density of know astrophysical sources such as FR I/II galaxies should be useful, as have been discussed for steady UHECR sources \citep[e.g.,][]{Takami2009Aph30p306}. A recent constraint on the apparent source number density is $\sim 10^{-4}$ Mpc$^{-3}$ above $\sim 6 \times 10^{19}$ eV \citep{Takami2009Aph30p306,Cuoco2009ApJ702p825}. This value can be regarded as $n_s(E)$ at $E \sim 6 \times 10^{19}$ eV because of the steep spectrum observed at around this energy with the spectral index of $\sim 4.3$~\citep{Abraham2010PhysLettB685p239}. Thus, for instance, FR II galaxies, which has the local number density of $\sim 3 \times 10^{-8}$ Mpc$^{-3}$~\citep{Woltjer1990agnconf}, and flat-spectrum radio quasars which correspond to on-axis FR II galaxies in the framework of the unification scenario of radio galaxies~\citep{Urry1995PASP107p803}, seem too rare as UHECR proton sources.

\begin{figure}
\includegraphics[clip,width=\linewidth]{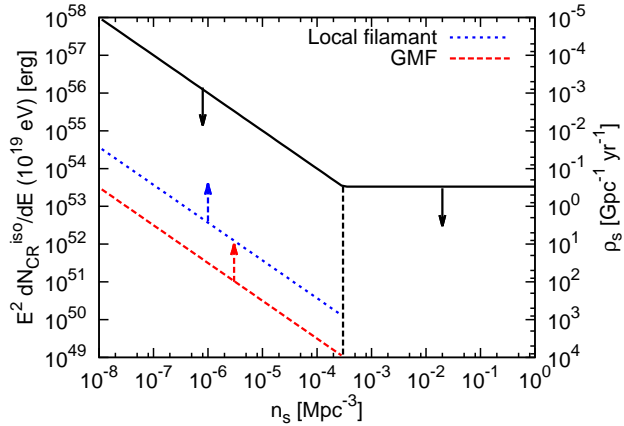}
\caption{Same as Figure~\ref{fig:einput}, but the possible case that a magnetic field in a local filament contributes to the propagation of UHE protons if UHECR sources are embedded in filamentary structures ({\it blue}). For reference, a limit when only the GMF contributes to the propagation, i.e., the effect of a magnetic structure around sources is negligible, is shown ({\it red}), which is regarded as a conservative case discussed in \cite{Murase2008ApJ690L14}. The black dashed line and solid line are the same as in Figure~\ref{fig:einput}. Note that, for the cases where sources are embedded in clusters of galaxies, constraints on $\rho_s$ and $\tilde{\mathcal{E}}^{\rm iso}_{\rm CR}$ given in Figure~\ref{fig:einput} are more relevant.} 
\label{fig:einput2}
\end{figure}

In this study, we adopt simple three zone models of magnetic fields (the GMF, a structured EGMF around sources, and the void EGMF), but modeling of the EGMFs have large uncertainty at present. Cosmological structure formation simulations have indicated that cosmic magnetic fields have complex web structures following matter distribution \citep{Sigl2003PRD68p043002,Dolag2005JCAP01p009,Das2008ApJ682p29}. At present, the volume filling fraction of relatively strong ($\gtrsim 10$ nG) magnetic fields, i.e., filaments and clusters, highly depends on models and methods (see Figure 9 of \cite{Kotera2011arXiv11014256}). Also, the spectrum of the magnetic fields is crucial for the propagation of UHECRs, though it is not certain. An observational constraint of the magnetic strength of EGMFs in filamentary structures is only $\lesssim 0.1 \mu$G \citep{Xu2006ApJ637p19} and there is no direct observational implication on the direction and coherent length of the magnetic fields. While a turbulent magnetic field with the Kolmogorov spectrum with $B_{\rm f} = 10$ nG is simply adopted in this work, several numerical simulations have indicated the existence of large-scale fields in filamentary structures~\citep[e.g.,][]{Bruggen2005ApJ631L21}. Such a coherent component deflects the trajectories of UHECRs more efficiently and makes longer time-delay compared to a turbulent component. In this case the time spread of a UHECR burst is smaller than the time-delay because particles similarly propagate due to the large-scale component.

The complex web of EGMFs implies that structured EGMFs on the way from sources to the Milky Way outside EGMFs embedding the sources could also play a significant role in the total deflection and time sperad of UHECRs, but these do not affect our constraints derived in the previous section in the scope of our EGMF models. Possible effects of such EGMFs are making additional deflection and time spread of UHECRs. As long as the additional deflection is small enough, the time spread by the EGMFs embedding the sources provides conservative constraints. In order to see the effects, we begin from evaluating the probability that UHECRs encounter a magnetic structure on the way to the Earth during propagation, which can be calculated from the number density, $n_{\rm x}$, and cross-sectional area, $A_{\rm x}$, of the structure x. The number density, $n_{\rm x} \approx f_{\rm x} / V_{\rm x}$, where $V_{\rm x}$ is the volume of the structure x, is $n_{\rm f} \approx 3 \times 10^{-5}$ Mpc$^{-3}$ and $n_{\rm c} \approx 10^{-6}$ Mpc$^{-3}$, respectively. The cross-sectional areas are $A_{\rm f} \approx 2^2 \times 25 = 100$ Mpc$^2$ and $A_{\rm c} \approx \pi \times 3^2 = 9 \pi$ Mpc$^2$, so the encounter probability is 0.24 and $2 \times 10^{-3}$ for 75 Mpc propagation in the cases of filaments and clusters, respectively. Thus, whereas it is unlikely for UHECRs to encounter clusters during their propagation, some of UHECRs penetrate into a filamentary structure once.  But their typical deviation angle of protons with energies above $10^{20}$~eV is still smaller than the considered value of $\psi = 5^{\circ}$ (see Figure~\ref{fig:def200}), so that such intervening EGMFs do not affect our conservative results. Stronger constraints on $\rho_s$ and $\tilde{\mathcal{E}}_{\rm CR}^{\rm iso}$ from additional time spread can be expected especially if UHECRs must pass the local magnetized structure, as discussed afterwards.

One might expect the situation where effects of structured EGMFs are even more prominent. We could misperceive the positions of UHECR sources from their arrival directions if UHECRs propagate selectively along the magnetic web structure and/or are strongly scattered off by the structured EGMFs~\citep{Kotera2008PRD77p123003,Ryu2010ApJ710p1422}. When such effects are large in the Universe, taking larger values of $\psi$ will be more realistic. Also, as recently suggested, the structured EGMFs may increase the probability that we observe UHECRs from transient sources thanks to significant time spread longer than the intrinsic burst duration of the sources~\citep{Kalli2011AA528p109}.

The Milky Way is thought to be located in a dense region in local Universe. Although magnetic fields in the local structure is highly uncertain, it could be another unavoidable EGMF for UHECRs arriving at the Earth. A constrained simulation has shown that the Milky Way belongs to a filamentary structure adjacent to the local Supercluster \citep{Klypin2003ApJ596p19}.~\footnote{A larger local magnetic field with $\sim 0.1~\mu \rm G$ may be possible if the Milky Way is located in the edge of the Virgo cluster \citep{Blasi1999PRD59p023001}.  Such a strong local magnetic field may be relevant as well as the void EGMF, especially when the strong effective magnetic field is required as in the scenario of high-luminosity GRBs~\citep{Murase2008ApJ690L14}.} Assuming our filament model for the local magnetic structure, this EGMF is dominant in $\tau_{\rm min}(E)$ in the case where sources are located in filamentary structures because the time profile spread should be estimated as $\sigma_{\rm f}(E)$ instead of $\tau_{\rm f}(E)$. Figure~\ref{fig:einput2} shows possible constraints on $\rho_s$ and $\tilde{\mathcal{E}}_{\rm CR}^{\rm iso}$ in the case where the local filament is included. Conservative constraints by the GMF, which is discussed in \cite{Murase2008ApJ690L14}, is also shown. For a filamentary structure between a source and the local environment discussed above, the time profile spread can also be conservatively estimated by $\sigma_{\rm f}(E)$, and therefore the same constraints as the case of the local filament are applied even if UHECRs pass several filamentary structures. In the cases that sources are embedded in clusters of galaxies, constraints are unchanged because the time spread of UHECRs by EGMFs in the clusters of galaxies is much larger. The discussions in Section~\ref{results}, where only EGMFs surrounding the sources are considered, provide conservative constraints. If the volume filling fraction of structured EGMFs and the nature of local magnetic environments are understood, we would be able to obtain better constraints.

As seen above, it is obviously crucial to reduce the uncertainty of the EGMFs both theoretically and observationally in future to understand the properties of transient UHECR sources. In order to get better knowledge on the EGMFs, Faraday rotation surveys by future experiments such as Square Kilometer Array\footnote{http://www.skatelescope.org/} are useful. Detection of TeV synchrotron pair echo/halo emission produced by UHE gamma-ray bursts/flares also may allow us to probe structured EGMFs with $\gtrsim 10$~nG~\citep{mur11}. Future numerical simulations on structured EGMFs can also give us more insight into consequences to transient UHECR source population.

Throughout this paper we have focused on protons. Even though heavy nuclei are considered, the discussions in this paper are in principle applicable. A basic difference between protons and heavy nuclei is electric charge, i.e., nuclei suffer from deflections $Z$ times as large as protons and the time spread roughly $Z^2$ times as large as protons. If structured EGMFs only in the vicinity of UHECR sources affect the propagation of nuclei, the discussions are valid because the deviation scale, $\psi$, reflects not the deflection angles of nuclei in the magnetic structures but a viewing angle of the structures. However, the GMF and a magnetic field in the Local Group would become seriously important. For the GMF, since the deflections of nuclei are too large at energies less than $10^{20}$ eV, we should focus on nuclei with energies much higher than $10^{20}$~eV. The deflection angles of nuclei could be the order of tens degree above a few times $10^{20}$ eV (e.g., \cite{Giacinti2010JCAP08p036}). Thus, even considering the GMF, the discussions can in principle possible with larger values of $\psi$ (e.g., $\psi \sim 30^{\circ}$). It is very uncertain that such discussions are possible even when an EGMF in local Group is taken into account because of more uncertainty. Future studies on effects of the EGMF in local Group will be necessary.

Although we have focused on an extragalactic component of UHECRs, possible Galactic sources, including GRBs~\citep{le93}, hypernovae~\citep{bud+07}, fast-rotating neutron stars~\citep{Blasi2000ApJ533L123} and magnetars, also could contribute to observed UHECRs.  In general, Galactic source scenarios have difficulty to reproduce anisotropy in the arrival distribution of UHECRs~\citep[e.g.,][]{Pohl2011ApJ742p114}. Recently \cite{Calvez2011PRL105p091101} suggested that Galactic sources would cause the gradual change of UHECR composition reported by the PAO and therefore also contribute to the UHECR flux around the ankle without inconsistency with observed anisotropy.  However, this scenario does not affect our discussions, since it still requires an extragalactic component of UHECRs above $3 \times 10^{19}$~eV, which we are interested in.

Although we have considered how we can reveal transient sources with observations of charged cosmic rays, for transient sources, the multi-messenger approach is definitely relevant, i.e., high-energy neutrino observations~\citep[e.g.,][]{wb97,rm98}, GeV-TeV gamma-rays~\citep[e.g.,][]{ad03,Dermer2009NJPh11p5016,mur11}, and UHE photons~\citep{Murase2009PRL103p081102} are helpful for identifying the sources. 

\section{Summary} \label{summary}

We studied the propagation of UHECRs with analytical models of structured EGMFs, and demonstrated the effects of the structured fields on revealing the properties of transient UHECR sources by UHECR experiments. While the void EGMF seems difficult to probe, structured EGMFs may be measured by future radio observations or revealed by dedicated numerical simulations. Then, UHECR experiments with large exposures may allow us to constrain transient UHECR source population by comparing the derived properties, i.e., $\rho_s$ and $\tilde{\mathcal{E}}_{\rm CR}^{\rm iso}$, with those of known astrophysical transients such as GRBs, AGN flares and magnetar generation (see table 4 of \cite{Murase2008ApJ690L14} for generation rates). We demonstrated that the energy-dependence of the apparent source number density $n_s(E)$ is a crucial hint of transient sources, where, as we argued, observations above ${10}^{20}$~eV are desirable due to the possible multiple-burst contamination at lower energies.  Once this transient feature is identified, the rate of UHECR bursts $\rho_s$ and energy input per burst $\tilde{\mathcal{E}}_{\rm CR}^{\rm iso}$ can be limited, given good knowledge on the structured EGMFs.

Finally, based on the discussions in this paper, we suggest a strategy to identify transient UHECR source population by future UHECR experiments with large exposures: 
\begin{enumerate}
\item{to estimate $n_s(E)$ from high event statistics and to see whether UHECR sources are transient from the energy dependence of $n_s(E)$,}
\item{to calculate the time profile spread of UHECRs from bursting or flaring sources, given good knowledge on the structured EGMFs from observations and/or simulations}
\item{to constrain $\rho_s$ and $\tilde{\mathcal{E}}_{\rm CR}^{\rm iso}$ from the above pieces of information}
\item{to discuss implications for known astrophysical candidates of UHECR sources via comparison with the properties of these transients.}
\end{enumerate}

\begin{acknowledgements}
We thank to P.~L.~Biermann, C.~D.~Dermer, and S.~Inoue for useful comments and discussions. We also grateful to the anonymous referee. The work of K.M. is supported by a Grant-in-Aid from Japan Society for the Promotion of Science (JSPS) and the Center of Cosmology and AstroParticle Physics (CCAPP). 
\end{acknowledgements}


\begin{thebibliography}{90}
\expandafter\ifx\csname natexlab\endcsname\relax\def\natexlab#1{#1}\fi

\bibitem[{{Abbasi} {et~al.}(2010)}]{Abbasi2010PRL104p161101}
{Abbasi}, R.~U., {et~al.} 2010, \prl, 104, 161101

\bibitem[{Abraham {et~al.}(2007)}]{Abraham2007Sci318p938}
Abraham, J., {et~al.} 2007, Science, 318, 938

\bibitem[{{Abraham} {et~al.}(2010{\natexlab{a}}){Abraham}, {Abreu}, {Aglietta},
  {Ahn}, {Allard}, {Allekotte}, {Allen}, {Alvarez-Mu{\~n}iz}, {Ambrosio},
  {Anchordoqui}, \& et~al.}]{Abraham2010PRL104p091101}
{Abraham}, J., {et~al.} 2010{\natexlab{a}}, \prl, 104, 091101

\bibitem[{{Abraham} {et~al.}(2010{\natexlab{b}}){Abraham}, {Abreu}, {Aglietta},
  {Ahn}, {Allard}, {Allen}, {Alvarez-Mu{\~n}iz}, {Ambrosio}, {Anchordoqui},
  {Andringa}, \& et~al.}]{Abraham2010PhysLettB685p239}
---. 2010{\natexlab{b}}, Phys.~Lett.~B, 685, 239

\bibitem[Abreu et al.(2011)]{abr11}
Abreu, P., et al. 2011, J. Cosmology Astropart. Phys., 06, 022 

\bibitem[Anchordoqui et al.(2011)]{anc11}
Anchordoqui, L.~A., et al. 2011, \prd, 84, 067301

\bibitem[{Alvarez-Mu\~niz {et~al.}(2002)Alvarez-Mu\~niz, Engel, \&
  Stanev}]{AlvarezMuniz:2001vf}
Alvarez-Mu\~niz, J., Engel, R., \& Stanev, T. 2002, \apj, 572, 185

\bibitem[{Arons(2003)}]{Arons2003ApJ589p871}
Arons, J. 2003, \apj, 589, 871

\bibitem[Atoyan \& Dermer (2003)]{ad03}
Atoyan, A., \& Dermer, C.~D. 2003, \apj, 586, 79

\bibitem[{Berezinsky {et~al.}(2006)Berezinsky, Gazizov, \&
  Grigorieva}]{Berezinsky:2002nc}
Berezinsky, V., Gazizov, A.~Z., \& Grigorieva, S.~I. 2006, \prd, 74, 043005

\bibitem[{{Biermann} \& {Strittmatter}(1987)}]{Biermann1987ApJ322p643}
{Biermann}, P.~L., \& {Strittmatter}, P.~A. 1987, \apj, 322, 643

\bibitem[Blandford (2000)]{bla00}
Blandford, R.~D. 2000, \physscr, T85, 191

\bibitem[{{Blandford} {et~al.}(1990){Blandford}, {Netzer}, {Woltjer},
  {Courvoisier}, \& {Mayor}}]{Woltjer1990agnconf}
{Blandford}, R.~D., {Netzer}, H., {Woltjer}, L., {Courvoisier}, T.~J.-L., \&
  {Mayor}, M., eds. 1990, {Active Galactic Nuclei}

\bibitem[{{Blasi} {et~al.}(1999){Blasi}, \& Olinto}]{Blasi1999PRD59p023001}
{Blasi}, P., \& {Olinto}, A.~V. 1999, \prd, 59, 023001

\bibitem[{{Blasi} {et~al.}(1999){Blasi}, {Burles}, \&
  {Olinto}}]{Blasi1999ApJ514L79}
{Blasi}, P., {Burles}, S., \& {Olinto}, A.~V. 1999, \apjl, 514,
  L79

\bibitem[{Blasi {et al.}(2000)Blasi, Epstein, \& Olinto}]{Blasi2000ApJ533L123}
{Blasi}, P., {Epstein}, R.~I., \& {Olinto}, A.~V. 2000, \apj, 533, L123

\bibitem[{{Bl{\"u}mer} \& {the Pierre Auger
  Collaboration}(2010)}]{Bluemer2010NJPh12c5001}
{Bl{\"u}mer}, J., \& {the Pierre Auger Collaboration}. 2010, New Journal of
  Phys., 12, 035001

\bibitem[{{Br{\"u}ggen} {et~al.}(2005){Br{\"u}ggen}, {Ruszkowski},
  {Simionescu}, {Hoeft}, \& {Dalla Vecchia}}]{Bruggen2005ApJ631L21}
{Br{\"u}ggen}, M., {Ruszkowski}, M., {Simionescu}, A., {Hoeft}, M., \& {Dalla
  Vecchia}, C. 2005, \apjl, 631, L21

\bibitem[Budnik et al.(2007)]{bud+07}
Budnik, R., Katz, B., MacFayden, A., \& Waxman, E. 2007, ApJ, 673, 928

\bibitem[{Calvez {et~al.}(2010)Calvez, Kusenko, \& Nagataki}]{Calvez2011PRL105p091101}
Calvez, A., Kusenko, A., \& Nagataki, S. 2010, \prl, 105, 091101

\bibitem[{Chodorowski {et~al.}(1992)Chodorowski, Zdziarski, \&
  Sikora}]{Chodorowski1992ApJ400p181}
Chodorowski, M.~J., Zdziarski, A.~A., \& Sikora, M. 1992, \apj, 400, 181

\bibitem[{{Cuoco} {et~al.}(2009){Cuoco}, {Hannestad}, {Haugb{\o}lle},
  {Kachelrie{\ss}}, \& {Serpico}}]{Cuoco2009ApJ702p825}
{Cuoco}, A., {Hannestad}, S., {Haugb{\o}lle}, T., {Kachelrie{\ss}}, M., \&
  {Serpico}, P.~D. 2009, \apj, 702, 825

\bibitem[{Das {et~al.}(2008)Das, Kang, Ryu, \& Cho}]{Das2008ApJ682p29}
Das, S., Kang, H., Ryu, D., \& Cho, J. 2008, \apj, 682, 29

\bibitem[{{de Marco} {et~al.}(2006){de Marco}, {Hansen}, {Stanev}, \&
  {Blasi}}]{DeMarco2006PhRvD73p043004}
{de Marco}, D., {Hansen}, P., {Stanev}, T., \& {Blasi}, P. 2006, \prd, 73,
  043004

\bibitem[{{Dermer} {et~al.}(2011){Dermer}, {Cavadini}, {Razzaque}, {Finke},
  {Chiang}, \& {Lott}}]{Dermer2011ApJ733L21}
{Dermer}, C.~D., {Cavadini}, M., {Razzaque}, S., {Finke}, J.~D., {Chiang}, J.,
  \& {Lott}, B. 2011, \apjl, 733, L21

\bibitem[{{Dermer} {et~al.}(2009){Dermer}, {Razzaque}, {Finke}, \&
  {Atoyan}}]{Dermer2009NJPh11p5016}
{Dermer}, C.~D., {Razzaque}, S., {Finke}, J.~D., \& {Atoyan}, A. 2009, New
  Journal of Phys., 11, 065016

\bibitem[{Dolag {et~al.}(2005)Dolag, Grasso, Springel, \&
  Tkachev}]{Dolag2005JCAP01p009}
Dolag, K., Grasso, D., Springel, V., \& Tkachev, I. 2005, J. Cosmology Astropart. Phys., 0501, 009

\bibitem[{{Dolag} {et~al.}(2011){Dolag}, {Kachelriess}, {Ostapchenko}, \&
  {Tom{\`a}s}}]{Dolag2011ApJ727L4}
{Dolag}, K., {Kachelriess}, M., {Ostapchenko}, S., \& {Tom{\`a}s}, R. 2011,
  \apjl, 727, L4

\bibitem[{{Ebisuzaki} {et~al.}(2008){Ebisuzaki}, {Uehara}, {Ohmori}, {Kawai},
  {Kawasaki}, {Sato}, {Takizawa}, {Bertaina}, {Kajino}, {Sawabe}, {Inoue},
  {Sasaki}, {Sakata}, {Yamamoto}, {Nagano}, {Inoue}, {Shibata}, {Sakaki},
  {Uchihori}, {Takahashi}, {Shimizu}, {Arai}, {Kurihara}, {Fujimoto},
  {Yoshida}, {Mizumoto}, {Inoue}, {Asano}, {Sugiyama}, {Watanabe}, {Ikeda},
  {Suzuki}, {Imamura}, {Yano}, {Murakami}, {Yonetoku}, {Itow}, {Taguchi},
  {Nagata}, {Nagataki}, {Abe}, {Tajima}, {Adams}, {Mitchell}, {Christl},
  {Watts}, {English}, {Takahashi}, {Pitalo}, {Hadaway}, {Geary}, {Readon},
  {Crawford}, {Pennypacker}, {Arisaka}, {Cline}, {Gorodetsky}, {Salin},
  {Patzark}, {Maurissen}, \& {Valentin}}]{Ebisuzaki2008NuPhS175p237}
{Ebisuzaki}, T., {et~al.} 2008, Nucl. Phys. B Proceedings Supplements, 175,
  237

\bibitem[{Farrar \& Gruzinov(2009)}]{Farrar:2008ex}
Farrar, G.~R., \& Gruzinov, A. 2009, \apj, 693, 329

\bibitem[{{Giacinti} {et~al.}(2010){Giacinti}, {Kachelrie{\ss}}, {Semikoz}, \&
  {Sigl}}]{Giacinti2010JCAP08p036}
{Giacinti}, G., {Kachelrie{\ss}}, M., {Semikoz}, D.~V., \& {Sigl}, G. 2010, J.
  Cosmo. Astropart. Phys., 8, 36

\bibitem[{Greisen(1966)}]{Greisen1966PRL16p748}
Greisen, K. 1966, \prl, 16, 748

\bibitem[Gorbunov et al.(2008)]{gor+08}
Gorbunov, D., et al. 2008, JETP Lett., 87, 461

\bibitem[{Hillas(1984)}]{Hillas1984ARAA22p425}
Hillas, A.~M. 1984, \araa, 22, 425

\bibitem[{Inoue {et~al.}(2007)Inoue, Sigl, Miniati, \&
  Armengaud}]{Inoue:2007kn}
Inoue, S., Sigl, G., Miniati, F., \& Armengaud, E. 2007, arXiv:astro-ph/0701167

\bibitem[Jedamzik et al.(2000)]{Jedamzik2000PRL85p700}
Jedamzik, K., Katalini{\'c}, V., \& Olinto, A.~V., \prl, 85, 700

\bibitem[{Kachelriess \& Semikoz(2006)}]{Kachelriess:2005xh}
Kachelriess, M., \& Semikoz, D.~V. 2006, Phys. Lett. B, 634, 143

\bibitem[{{Kalli} {et~al.}(2011){Kalli}, {Lemoine}, \&
  {Kotera}}]{Kalli2011AA528p109}
{Kalli}, S., {Lemoine}, M., \& {Kotera}, K. 2011, \aap, 528, A109

\bibitem[{{Kang} {et~al.}(1996){Kang}, {Ryu}, \& {Jones}}]{Kang1996ApJ456p422}
{Kang}, H., {Ryu}, D., \& {Jones}, T.~W. 1996, \apj, 456, 422

\bibitem[{{Klypin} {et~al.}(2003){Klypin}, {Hoffman}, {Kravtsov}, \&
  {Gottl{\"o}ber}}]{Klypin2003ApJ596p19}
{Klypin}, A., {Hoffman}, Y., {Kravtsov}, A.~V., \& {Gottl{\"o}ber}, S. 2003,
  \apj, 596, 19

\bibitem[{Kneiske {et~al.}(2004)Kneiske, Bretz, Mannheim, \&
  Hartmann}]{Kneiske2004AA413p807}
Kneiske, T.~M., Bretz, T., Mannheim, K., \& Hartmann, D.~H. 2004, \aap, 413, 807

\bibitem[{{Kotera}(2011)}]{Kotera2011PhRvD84p023002}
{Kotera}, K. 2011, \prd, 84, 023002

\bibitem[{Kotera \& Lemoine(2008)}]{Kotera2008PRD77p123003}
Kotera, K., \& Lemoine, M. 2008, \prd, 77, 123003

\bibitem[{{Kotera} \& {Olinto}(2011)}]{Kotera2011arXiv11014256}
{Kotera}, K., \& {Olinto}, A.~V. 2011, \araa, 29, 119

\bibitem[{Kronberg(1994)}]{Kronberg1994RepProgPhys57p325}
Kronberg, P.~P. 1994, Rept. Prog. Phys., 57, 325

\bibitem[Lemoine \& Waxman(2009)]{lw09}
Lemoine, M., \& Waxman, E. 2009, J. Cosmology Astropart. Phys., 11, 009

\bibitem[Levinson \& Eichler(1993)]{le93}
Levinson, A., \& Eichler, D. 1993, ApJ, 418, 386

\bibitem[{{Lin} \& {Mohr}(2007)}]{Lin2007ApJS170p71}
{Lin}, Y.-T., \& {Mohr}, J.~J. 2007, \apjs, 170, 71

\bibitem[{{Mazure} {et~al.}(1996){Mazure}, {Katgert}, {den Hartog}, {Biviano},
  {Dubath}, {Escalera}, {Focardi}, {Gerbal}, {Giuricin}, {Jones}, {Le Fevre},
  {Moles}, {Perea}, \& {Rhee}}]{Mazure1996AA310p31}
{Mazure}, A., {et~al.} 1996, \aap, 310, 31

\bibitem[{M$\ddot{\rm u}$cke {et~al.}(2000)M$\ddot{\rm u}$cke, Engel, Rachen,
  Protheroe, \& Stanev}]{Mucke1999CPC124p290}
M$\ddot{\rm u}$cke, A., Engel, R., Rachen, J.~P., Protheroe, R.~J., \& Stanev,
  T. 2000, Comput. Phys. Commun., 124, 290

\bibitem[{Miralda-Escude \& Waxman(1996)}]{MiraldaEscude1996ApJ462L59}
Miralda-Escude, J., \& Waxman, E. 1996, \apj, 462, L59

\bibitem[{{Murase}(2009)}]{Murase2009PRL103p081102}
{Murase}, K. 2009, \prl, 103, 081102

\bibitem[{{Murase}(2011)}]{mur11}
{Murase}, K. 2011, arXiv:1111.0936

\bibitem[{Murase \& Takami(2009)}]{Murase2008ApJ690L14}
Murase, K., \& Takami, H. 2009, \apjl, 690, L14

\bibitem[{{Murase} {et~al.}(2011){Murase}, {Dermer}, {Takami}, \&
  {Migliori}}]{2011arXiv1107.5576M}
{Murase}, K., {Dermer}, C.~D., {Takami}, H., \& {Migliori}, G. 2011, Arxiv: 1107.5576

\bibitem[{Murase {et~al.}(2006)Murase, Ioka, Nagataki, \&  Nakamura}]{Murase:2006mm}
Murase, K., Ioka, K., Nagataki, S., \& Nakamura, T. 2006, \apj, 651, L5

\bibitem[{{Murase} {et~al.}(2008a){Murase}, {Ioka}, {Nagataki}, \&  {Nakamura}}]{Murase2008PRD78p023005}
{Murase}, K., {Ioka}, K., {Nagataki}, S., \& {Nakamura}, T. 2008a, \prd, 78, 023005

\bibitem[{Murase {et~al.}(2009)Murase, Meszaros, \&  Zhang}]{Murase2009PRD79p103001}
Murase, K., Meszaros, P., \& Zhang, B. 2009, \prd, 79, 103001

\bibitem[Murase et al. (2008b)]{mur08b} 
Murase, K., Takahashi, K., Inoue, S., Ichiki, K., \& Nagataki, S.\ 2008b, \apjl, 686, L67 

\bibitem[{{Norman} {et~al.}(1995){Norman}, {Melrose}, \&
  {Achterberg}}]{Norman:1995aa}
{Norman}, C.~A., {Melrose}, D.~B., \& {Achterberg}, A. 1995, \apj, 454, 60

\bibitem[{{Padovani} \& {Urry}(1990)}]{Padovani1990ApJ356p75}
{Padovani}, P., \& {Urry}, C.~M. 1990, \apj, 356, 75

\bibitem[{{Pohl} \& {Eichler}(2011)}]{Pohl2011ApJ742p114}
{Pohl}, M., \& {Eichler}, D. 2011, \apj, 742, 114

\bibitem[{Pe'er {et~al.}(2009)Pe'er, Murase, \&
  Meszaros}]{Pe'er2009PRD80p123018}
Pe'er, A., Murase, K., \& Meszaros, P. 2009, \prd, 80, 123018

\bibitem[Plaga (1995)]{pla95}
Plaga, R. 1995, \nat, 374, 430

\bibitem[{{Porter} {et~al.}(2008){Porter}, {Raychaudhury}, {Pimbblet}, \&
  {Drinkwater}}]{Porter2008MNRAS388p1152}
{Porter}, S.~C., {Raychaudhury}, S., {Pimbblet}, K.~A., \& {Drinkwater}, M.~J.
  2008, \mnras, 388, 1152

\bibitem[Rachen \& M\'esz\'aros (1998)]{rm98}
Rachen, J.~P., \& M\'esz\'aros, P. 1998, \prd, 58, 123005

\bibitem[{{Rordorf} {et~al.}(2004){Rordorf}, {Grasso}, \&
  {Dolag}}]{Rordorf2004APh22p167}
{Rordorf}, C., {Grasso}, D., \& {Dolag}, K. 2004, Astropart. Phys., 22,
  167

\bibitem[{{Ryu} {et~al.}(2010){Ryu}, {Das}, \& {Kang}}]{Ryu2010ApJ710p1422}
{Ryu}, D., {Das}, S., \& {Kang}, H. 2010, \apj, 710, 1422

\bibitem[{{Ryu} {et~al.}(1998){Ryu}, {Kang}, \& {Biermann}}]{Ryu1998AA335p19}
{Ryu}, D., {Kang}, H., \& {Biermann}, P.~L. 1998, \aap, 335, 19

\bibitem[{Ryu {et~al.}(2008)Ryu, Kang, Cho, \& Das}]{Ryu2008Science320p909}
Ryu, D., Kang, H., Cho, J., \& Das, S. 2008, Science, 320, 909

\bibitem[{Sigl {et~al.}(2003)Sigl, Miniati, \& Ensslin}]{Sigl2003PRD68p043002}
Sigl, G., Miniati, F., \& Ensslin, T.~A. 2003, \prd, 68, 043002

\bibitem[{Sigl {et~al.}(2004)Sigl, Miniati, \& Ensslin}]{Sigl2004PRD70p043007}
---. 2004, \prd, 70, 043007

\bibitem[{{Silva} {et~al.}(1998){Silva}, {Granato}, {Bressan}, \&
  {Danese}}]{Silva1998ApJ509p103}
{Silva}, L., {Granato}, G.~L., {Bressan}, A., \& {Danese}, L. 1998, \apj, 509, 103

\bibitem[{Takahara(1990)}]{Takahara:1990he}
Takahara, F. 1990, Prog. Theor. Phys., 83, 1071

\bibitem[{{Takahashi} {et~al.}(2011){Takahashi}, {Mori}, {Ichiki}, \&
  {Inoue}}]{Takahashi2011arXiv11033835}
{Takahashi}, K., {Mori}, M., {Ichiki}, K., \& {Inoue}, S. 2011, ArXiv: 1103.3835

\bibitem[{{Takami} \& {Horiuchi}(2011)}]{Takami2011APh34p749}
{Takami}, H., \& {Horiuchi}, S. 2011, Astropart. Phys., 34, 749

\bibitem[{{Takami} {et~al.}(2009){Takami}, {Nishimichi}, \&
  {Sato}}]{2009arXiv0910.2765T}
{Takami}, H., {Nishimichi}, T., \& {Sato}, K. 2009, ArXiv: 0910.2765

\bibitem[{Takami {et~al.}(2009)Takami, Nishimichi, Yahata, \&
  Sato}]{Takami2008JCAP06p031}
Takami, H., Nishimichi, T., Yahata, K., \& Sato, K. 2009, J. Cosmology. Astropart. Phys., 0906, 031

\bibitem[{Takami \& Sato(2008)}]{Takami:2007kq}
Takami, H., \& Sato, K. 2008, \apj, 681, 1279

\bibitem[{Takami \& Sato(2008)}]{Takami2007Aph28p529}
---. 2008, Astropart. Phys., 28, 529

\bibitem[{Takami \& Sato(2009)}]{Takami2009Aph30p306}
---. 2009, Astropart. Phys., 30, 306

\bibitem[{{Takami} \& {Sato}(2010)}]{Takami2010ApJ724p1456}
{Takami}, H., \& {Sato}, K. 2010, \apj, 724, 1456

\bibitem[{Takami {et~al.}(2006)Takami, Yoshiguchi, \&
  Sato}]{Takami2006ApJ639p803}
Takami, H., Yoshiguchi, H., \& Sato, K. 2006, \apj, 639, 803

\bibitem[{{Urry} \& {Padovani}(1995)}]{Urry1995PASP107p803}
{Urry}, C.~M., \& {Padovani}, P. 1995, \pasp, 107, 803

\bibitem[{Vietri(1995)}]{Vietri1995ApJ453p883}
Vietri, M. 1995, \apj, 453, 883

\bibitem[Wang et al.(2008)]{Wang2008ApJ677p432}
Wang, X.~-Y., Razzaque, S., \& M{\'e}sz{\'a}ros, P., \apj, 677, 432

\bibitem[{Waxman(1995)}]{Waxman:1995vg}
Waxman, E. 1995, \prl, 75, 386

\bibitem[{Waxman(2004)}]{Waxman:2003uj}
---. 2004, Pramana, 62, 483

\bibitem[Waxman \& Bahcall (1997)]{wb97}
Waxman, E., \& Bahcall, J. 1997, \prl, 78, 2292

\bibitem[{Waxman \& Bahcall(1999)}]{Waxman:1998yy}
Waxman, E., \& Bahcall, J.~N. 1999, \prd, 59, 023002

\bibitem[{{Waxman} \& {Miralda-Escude}(1996)}]{Waxman1996ApJ472L89}
{Waxman}, E., \& {Miralda-Escude}, J. 1996, \apj, 472, L89

\bibitem[Wilk \& Wlodarczyk(2011)]{ww11}
Wilk, G., \& Wlodarczyk, Z. 2011, J. Phys. G, 38, 085201  

\bibitem[{{Xu} {et~al.}(2006){Xu}, {Kronberg}, {Habib}, \&
  {Dufton}}]{Xu2006ApJ637p19}
{Xu}, Y., {Kronberg}, P.~P., {Habib}, S., \& {Dufton}, Q.~W. 2006, \apj, 637, 19

\bibitem[Yamazaki et al.(2010)]{Yamazaki2010PRD81p023008}
Yamazaki, D.~G., Ichiki, K., Kajino, T., \& Mathews, G.~J., \prd, 81, 023008

\bibitem[{Yoshiguchi {et~al.}(2003)Yoshiguchi, Nagataki, Tsubaki, \&
  Sato}]{Yoshiguchi2003ApJ586p1211}
Yoshiguchi, H., Nagataki, S., Tsubaki, S., \& Sato, K. 2003, \apj,
  586, 1211

\bibitem[{Zatsepin \& Kuz'min(1966)}]{Zatsepin1966JETP4L78}
Zatsepin, G.~T., \& Kuz'min, V.~A. 1966, JETP Lett., 4, 78

\bibitem[Zaw et al.(2009)]{zaw09}
Zaw, I., Farrar, G.~R., \& Greene, J. 2009, \apj, 696, 1218
\end{thebibliography}
\end{document}